\documentstyle[12pt,prl,aps,psfig]{revtex}
\begin{document}
\draft
\title{Perturbation Approach \\ for a Solid-State Quantum Computation}
\author{G.P. Berman$^\dag$, D.I. Kamenev$^{\dag}$,
and V.I. Tsifrinovich$^\ddag$}
\address{$^\dag$ Theoretical Division and CNLS, Los Alamos National Laboratory,
Los Alamos, NM 87545}
\address{$^\ddag$ IDS Department, Polytechnic University, Six Metrotech
Center, Brooklyn, New York 11201}
\maketitle
\vspace{5mm}
\begin{abstract}
The dynamics of the nuclear-spin quantum computer with large number
($L=1000$) of qubits is considered using a perturbation approach, based on approximate diagonalization of exponentially large sparse matrices. Small
parameters are introduced and used to compute the error in
implementation of entanglement between remote qubits, by
applying a sequence of resonant radio-frequency pulses. The results of the perturbation theory are tested using exact numerical solutions for small number of qubits.
\end{abstract}

\section{Introduction}

Different solid-state quantum systems are
considered now as candidates for quantum bits (qubits). They include:
nuclear spins \cite{10,11,12}, electron spins \cite{13,14,15},
electron in a quantum dot \cite{16}, and Josephson junctions \cite{17,18}.
For the most effective quantum information
processing the quantum computer should operate with large number
of qubits, $L$. For understanding how quantum computer works and for optimizing parameters of quantum computation it is necessary to model and simulate quantum logic operations and fragments of quantum algorizms. For numerical simulation of quantum computer dynamics 
one must solve a large set of $2^L$
linear differential equations on a long enough time-interval, or to diagonalize
many large matrices of the size $2^L\times 2^L$.
Hence, it is important to develop a consistent perturbation
theory for quantum computation
which allows one to predict the probability of correct implementation of 
quantum logic operations involving large number of qubits,
in the real physical systems.

In this paper we describe the procedure which allows one
to estimate the errors in implementation of quantum logic operations
in the one-dimensional solid-state system of nuclear spins. Our approach does not require exact
solution of quantum dynamical equations and direct
diagonalization of
large matrices. We assume that our quantum computer operates at zero
temperature, and the error is generated only as a result of
internal decoherence (non-resonant processes).
Our approach provides the tool for choosing optimal parameters for
operation of the scalable quantum computer with large number of qubits.

\section{Dynamics of the spin chain quantum computer}

Application of Ising spin systems for quantum
computations was first suggested in Ref.~\cite{ber3}. Today, similar  systems
are used in liquid nuclear magnetic resonance (NMR) quantum computation
with small number of qubits \cite{chuang}.
The register (a 1D chain of $N$ identical nuclear spins)
is placed in a magnetic field,
\begin{equation}
\label{1}
{\rm\bf B}^{(n)}(z,\,t)=
\left(b^{(n)}_\perp\cos\left[\nu^{(n)} t+\varphi^{(n)}\right],
-b^{(n)}_\perp\sin\left[\nu^{(n)} t+\varphi^{(n)}\right], B^z(z)\right),
\end{equation}
where $t^{(n)}\le t\le t^{(n+1)}$, $n=1,...,M$,
$B^z(z)$ is a slightly non-uniform magnetic
field oriented in the positive $z$-direction,
$b^{(n)}_\perp$, $\nu^{(n)}$ and $\varphi^{(n)}$
are, respectively, the amplitude, the frequency, and the initial phase
of the circular polarized (in the $x-y$ plane) magnetic field. This
magnetic field has the form of rectangular pulses of the length (time
duration) $\tau^{(n)}=t^{(n+1)}-t^{(n)}$. The total number of pulses
which is required to perform a given quantum computation (protocol) is
$M$. The chain of spins makes an angle $\theta$, where
$\cos(\theta)=1/\sqrt 3$, with
the direction of the magnetic field $B^z(z)$ to suppress the dipole
interaction between the spins \cite{1}.

The Hamiltonian of
the spin chain in the magnetic field is,
$$
H^{(n)}=-\sum_{k=0}^{L-1}\omega_kI_k^z-2J\sum_{k=0}^{L-1}I_k^zI_{k+1}^z-
$$
\begin{equation}
\label{H}
\Theta^{(n)}(t)(\Omega_n/2)\sum_{k=0}^{L-1}\left\{
I_k^-\exp\left[-i\left(\nu^{(n)} t+\varphi^{(n)}\right)\right]+
I_k^+\exp\left[i\left(\nu^{(n)} t+\varphi^{(n)}\right)\right]\right\}=
H_0+V^{(n)}(t),
\end{equation}
where the index $k$ labels the spins in the chain,
$J$ is the Ising interaction constant,
$\Omega_n=\gamma b^{(n)}_\perp$ is the precession (Rabi) frequency,
$\gamma$ is the gyromagnetic ratio.
The function $\Theta^{(n)}(t)$ equals $1$ only during the $n$th pulse
and equals to zero otherwise.

In order to remove the time-dependence from the Hamiltonian (\ref{H})
we present the wave function, $\Psi(t)$, in the time-interval
of the $n$th pulse, in the laboratory system of coordinates in the form,
\begin{equation}
\label{Psi1}
\Psi(t)=
\exp\left[i(\nu^{(n)}t+\varphi^{(n)})\sum_{k=0}^{L-1}
I_k^z\right]\Psi_{rot}^{(n)}(t)=
\sum_pA_p(t)|p\rangle\exp(-i\chi_p^{(n)}t+\xi_p^{(n)}),
\end{equation}
where $\Psi_{rot}^{(n)}(t)$ is the wave function in a reference frame rotating with
the frequency, $\nu^{(n)}$,
$\chi_p^{(n)}=-(\nu^{(n)}/2)\sum_{k=0}^{L-1}\sigma_k^p$,
$\xi_p^{(n)}=\varphi^{(n)}\sum_{k=0}^{L-1}\sigma_k^p$,
$\sigma_k^p=-1$ if the $k$th spin of the state $|p\rangle$
is in the position
$|1\rangle$ and $\sigma_k^p=1$ if the $k$th spin is in the position
$|0\rangle$, $|p\rangle$ is the eigenfunction of the Hamiltonian $H_0$.
Below we take $\varphi^{(n)}=\xi_p^{(n)}=0$ for all $n$.

The dynamics during the $n$th pulse is described
by the following Schr\"odinger equation for the coefficients $A_p(t)$,
\begin{equation}
\label{Sch2}
i\dot A_p(t)=(E_p-\chi_p^{(n)})A_p(t)-\frac\Omega 2\sum_{p'}A_{p'}(t),
\end{equation}
where we put, $\hbar=1$, for the Planck constant,
and the sum is taken over the states $|p'\rangle$
connected by a single-spin transition with the state $|p\rangle$,
$E_p$ is the eigenvalue of the Hamiltonian $H_0$.
In the representation (\ref{Psi1}) each spin flip is accompanied
by a change of the phase of the wave function by the value $\pm\nu^{(n)} t$,
which compensates the time-dependent phase in the perturbation,
$V^{(n)}(t)$, in the Hamiltonian
(\ref{H}). In this case, the dynamics of the coefficients, $A_p(t)$, is
governed by the effective time-independent Hamiltonian, ${\cal H}^{(n)}$,
and Eq. (\ref{Sch2}) can be written in the form,
$i\dot A_p(t)={\cal H}^{(n)}_{pp'}A_{p'}(t)$.

The dynamics of the coefficients, $A_{p}(t)$, generated by the Hamiltonian,
${\cal H}^{(n)}$, can be computed using the eigenfunctions,
$A_{m}^{q\,(n)}$, and the eigenvalues, $e_q^{n}$, of this Hamiltonian by,
\begin{equation}
\label{dynamics}
A_m(t_n)=\sum_{m_0}A_{m_0}(t_{n-1})
\sum_{q}A_{m_0}^{q\,(n)}A_m^{q\,(n)}\exp(-ie_q^{n}\tau_n).
\end{equation}
Since the pulses of the protocol are different, we should operate
in the laboratory frame and make the transformation of the wave function
to the rotating frame
before each pulse, and the transformation to the laboratory frame after
each pulse.
If we write the wave function in the laboratory frame in the form,
\begin{equation}
\label{Psi}
\Psi(t)=\sum_pC_p(t)|p\rangle\exp(-iE_pt),
\end{equation}
then the coefficients, $C_p(t)$, in the laboratory frame are expressed
through the coefficients, $A_p(t)$, in the rotating frame as,
\begin{equation}
\label{AC}
C_p(t)=\exp\left(i{\cal E}_p^{(n)}t\right)A_p(t).
\end{equation}
Here ${\cal E}_p^{(n)}=E_p-\chi_p^{(n)}$ are the diagonal elements
of the Hamiltonian matrix ${\cal H}_{pp'}^{(n)}$, $t=t_{n-1}$ before
the $n$th pulse and $t=t_n=t_{n-1}+\tau_n$ after the $n$th pulse.

\section{The two-level approximation}
We explain below how
selective (resonant) transitions,
which realize a quantum logic gate, can be implemented in the system
described by the Hamiltonian (\ref{H}).
For this, we consider the structure of the effective
time-independent Hamiltonian matrix, ${\cal H}_{pp'}$,
(here and below we omit the upper index, $n$.)
All non-zero non-diagonal matrix elements are the same and equal
to $-\Omega/2$.
At $\Omega\ll\delta\omega$ the absolute values of the
diagonal elements in general case
are much larger than the absolute values of the
off-diagonal elements, and
the resonance is coded in the structure of the diagonal
elements of the Hamiltonian matrix, ${\cal H}_{pp'}$.

Suppose that we want to flip the $k$th spin in the chain.
To do this, we choose the frequency of the pulse to be equal
to the difference, $\nu=E_p-E_m$, between the energies of the states
which are related to each other by flip of the $k$th spin.
In this case the energy separation between the $p$th and the $m$th
diagonal elements of the matrix, ${\cal H}_{pp'}$,
related by the flip of the resonant $k$th spin
is much less than the energy separation
between the $p$th diagonal elements and diagonal
elements related to other states, which differ from the state
$|p\rangle$ by a flip of a non-resonant $k'$th
($k'\ne k$) spin.
In this situation, in some approximation, one can neglect the interaction of the $p$th state with
all states except for the state, $|m\rangle$, and the Hamiltonian matrix,
${\cal H}_{pp'}$, breaks up into $2^L/2$,
approximately independent $2\times 2$ matrices,
\begin{equation}
\label{2x2}
\pmatrix{{\cal E}_{l} & V  \cr
V & {\cal E}_{m}},
\end{equation}
where ${\cal E}_{p}={\cal E}_{m}+\Delta_{pm}$,
$|\Delta_{pm}|\sim J$ or $|\Delta_{pm}|=0$
is the detuning from the resonance, which depends on
the positions of $(k-1)$th and $(k+1)$ spins, $V=-\Omega/2$.

Suppose, for example,  that $L=5$ and the third spin ($k=3$) has
resonant ($|\Delta_{pm}|=0$) or near-resonant ($|\Delta_{pm}|\sim J$)
frequency (we start enumeration from the right spin with $k=0$).
Then, the block $2\times 2$ will be organized, for example,
by the following states: $|01010\rangle$ and
$|00010\rangle$;
$|01111\rangle$ and $|00111\rangle$; $|00001\rangle$ and $|01001\rangle$,
and so on. In order to find the state, $|m\rangle$, which forms a $2\times 2$
block with a definite state $|p\rangle$, one should flip
the resonant spin of the state $|p\rangle$. In other words,
positions of $N-1$ (non-resonant) spins of these states are equivalent,
while position of the resonant spin is different. This approximation can be
called the two-level approximation, since in this case we have
$2^{L}/2$ independent two-level systems.

We now obtain the solution in the two-level approximation.
The dynamics is given by Eq. (\ref{dynamics}). Since we deal only with
a single $2\times 2$ block of the matrix ${\cal H}_{pp'}$,
(but not with the whole matrix), the dynamics in this approximation
is generated only by the eigenstates of one block.

The eigenvalues
$e^{(0)}_q$, $e^{(0)}_Q$, and the eigenfunctions
of the $2\times 2$ matrix (\ref{2x2}) are
(we put $\Delta_{pm}=\Delta$),
\begin{equation}
\label{q01}
\hspace{-9mm}e^{(0)}_q={\cal E}_m+\frac\Delta 2-\frac\lambda 2, \qquad
\pmatrix{
A^{q\,(0)}_m \cr A^{q\,(0)}_p}=
{1\over\sqrt{(\lambda-\Delta)^2+\Omega^2}}
\pmatrix{\Omega\cr \lambda-\Delta},
\end{equation}
\begin{equation}
\label{q02}
e^{(0)}_Q={\cal E}_m+\frac\Delta 2+\frac\lambda 2, \qquad
\pmatrix{
A^{Q\,(0)}_m\cr A^{Q\,(0)}_p}=
{1\over\sqrt{(\lambda-\Delta)^2+\Omega^2}}
\pmatrix{
-(\lambda-\Delta)\cr \Omega},
\end{equation}
where $\lambda=\sqrt{\Omega^2+\Delta^2}$.
Suppose that before the pulse the system is in the state $|m\rangle$,
i.e. the conditions,
$$
C_m(t_0)=1,\qquad C_p(t_{0})=0,
$$
are satisfied. After the transformation (\ref{AC}) to the rotating frame
we obtain,
$$
A_m(t_{0})=\exp(-i{\cal E}_mt_{0})C_m(t_{0})
=\exp(-i{\cal E}_mt_{0}),\qquad A_p(t_{0})=0.
$$
The dynamics is given by Eq. (\ref{dynamics}), which in our case
takes the form:
$$
A_m(t)=
A_m(t_{0})\left[\left(A_m^{q\,(0)}\right)^2\exp\left(-ie^{(0)}_q\tau\right)+
\left(A_m^{Q\,(0)}\right)^2\exp\left(-ie^{(0)}_Q\tau\right)\right]=
$$
$$
{\exp\left[-i[{\cal E}_mt-(\Delta/2)\tau]\right]\over
\Omega^2+(\lambda-\Delta)^2}\left\{\Omega^2 e^{-i\lambda\tau/2}+
(\lambda-\Delta)^2e^{i\lambda\tau/2}\right\},
$$
where $t=t_0+\tau$.
Applying the back transformation,
$$
C_m(t)=\exp(i{\cal E}_mt)A_m(t),
$$
and taking the real and imaginary parts of the expression
in the curl brackets we obtain,
\begin{equation}
\label{C1}
C_m(t_0+\tau)=\left[\cos(\lambda\tau/2)+
i(\Delta/\lambda)\sin(\lambda\tau/2)\right]
\exp(-i\tau\Delta/2).
\end{equation}
For the amplitude, $A_p(t)$, we have,
$$
A_p(t)=A_m(t_{0})
\left[A_m^{q\,(0)}A_p^{q\,(0)}\exp\left(-ie^{(0)}_q\tau\right)+
A_m^{Q\,(0)}A_p^{Q\,(0)}\exp\left(-ie^{(0)}_Q\tau\right)\right]=
$$
$$
i{\Omega\over\lambda}
\exp\left\{-i\left[{\cal E}_mt-(\Delta/2)\tau\right]\right\}
\sin(\lambda\tau/2).
$$
Applying the back transformation,
$$
C_p(t)=\exp[i({\cal E}_m+\Delta)t]A_p(t),
$$
we obtain,
\begin{equation}
\label{C2}
C_p(t_0+\tau)=i(\Omega/\lambda)\sin(\lambda\tau/2)
\exp(it_0\Delta+i\tau\Delta/2),
\end{equation}

When $\Delta=0$ (the resonance case) and $\lambda\tau=\pi$ ($\pi$-pulse)
Eqs. (\ref{C1}) and (\ref{C2}) describe the complete transition from
the state $|m\rangle$ to the state $|p\rangle$. In the near-resonance case,
when $\Delta\ne 0$ the transition probability is
(here we again put the index $n$ indicating the pulse number),
\begin{equation}
\label{epsilon}
\varepsilon_n=(\Omega_n/\lambda_n)^2\sin^2(\lambda_n\tau_n/2).
\end{equation}
One can suppress the near-resonant transitions and to make
the probability, $\varepsilon_n$, equal to zero by
choosing the Rabi frequency in the form
(see $2\pi k$-method in Ref. \cite{1}),
\begin{equation}
\label{2pi_k}
\Omega^{(k)}_n=|\Delta_n|/\sqrt{4k^2-1}.
\end{equation}

The solutions (\ref{C1}) and (\ref{C2}) can also be derived
without transformation
to the rotating frame (see Ref. \cite{1}). However, as will be shown below,
our description allows us to introduce small parameters and to build
a consistent perturbation theory. Using our perturbation approach we will compute the dynamics up to different orders in small parameters, and will test
our approximate results using exact numerical solution for small number of qubits.
Below we apply the perturbation theory for
analysis of the quantum dynamics during implementation
of a simple quantum logic gate in the spin chain with
large number ($L=1000$) of qubits.

\section{Protocol for creation an entangled state \\
between remote qubits}
Here we schematically describe the protocol (the sequence of pulses)
which allows one to create
the entangled state for remote qubits in the system described by the
Hamiltonian~(\ref{H}). The initial state of the system is the ground state, $|00\dots00\rangle$.
The first pulse
in our protocol, described by the unitary operator $U_1$,
creates the superposition of the states $|00\dots00\rangle$ and
$|10\dots00\rangle$ from the ground state,
\begin{equation}
\label{U0}
U_1|00\dots00\rangle={1\over \sqrt 2}(|00\dots00\rangle+i|10\dots00\rangle).
\end{equation}
Other pulses create from this state the entangled state for remote qubits.
This procedure is described by the unitary operator, $U'$,
\begin{equation}
\label{U}
U'\frac 1{\sqrt 2}(|00\dots00\rangle+
i|10\dots00\rangle)=\frac 1{\sqrt 2}(e^{i\varphi_1}(|00\dots00\rangle+
e^{i\varphi_2}|10\dots 01\rangle),
\end{equation}
where $\varphi_1$ and $\varphi_2$ are known phases~\cite{1}.

Now, we describe the procedure for realization of the operator
$U=U'U_1$ by applying a sequence of resonant pulses.
Each pulse is described by the corresponding unitary operator, $U_n$,
where $n=1,\,2,\,\dots,\,2L-2$. (The total number of pulses
in our protocol is $M=2L-2$.)
The unitary operator of the whole protocol is a product
of the unitary operators of the individual pulses,
$U=U_{2L-2}U_{2L-4}\dots U_2U_1$.
The first pulse, described by the operator $U_1$,
is resonant to the transition between the states $|00\dots00\rangle$ and
$|10\dots00\rangle$. If we choose the duration of this pulse as
$\tau_1=\pi/(2\Omega_1)$ (a $\pi/2$-pulse) then from Eqs. (\ref{C1}) and
(\ref{C2}) we obtain Eq. (\ref{U0}).
In order to obtain the second term in the right-hand side of Eq. (\ref{U}),
we choose a sequence of resonant $\pi$-pulses which transforms
the state $|10\dots 00\rangle$ to the state
$|10\dots 01\rangle$ by the following scheme:
$
|1000\dots 0\rangle\rightarrow|1100\dots 0\rangle\rightarrow
|1110\dots 0\rangle\rightarrow
$
$
|1010\dots 0\rangle\rightarrow|1011\dots 0\rangle\rightarrow
|1001\dots 0\rangle\rightarrow
$
$
\dots\rightarrow|100\dots 11\rangle\rightarrow|100\dots 01\rangle.
$
The frequencies of pulses which realize this protocol are:
$\nu^{(2)}=\omega_{2L-2}$, $\nu^{(3)}=\omega_{2L-3}$,
$\nu^{(4)}=\omega_{2L-2}-2J$, $\nu^{(5)}=\omega_{2L-4}$,
\dots, $\nu^{(2L-3)}=\omega_0-J$, $\nu^{(2L-2)}=\omega_1$.
If we apply the same protocol to the ground state,
then with large probability the system will remain in this state
because all transitions are non-resonant to the ground state.

Since the values of the detuning are the same for all
pulses, $\Delta_n=\Delta=2J$ (except for the fourth
pulse, where $\Delta_4=4J$), in our calculations
we take the values of $\Omega_n$ to be the same,
$\Omega_n=\Omega$ ($n\ne 4$), and $\Omega_4=2\Omega$.
In this case, the probabilities of excitation of the
ground state (near-resonant transitions),
$\varepsilon_n$,
are independent of $n$: $\varepsilon_n=\varepsilon$,
since $\varepsilon_n$ depends only on the ratio
$|\Delta_n/\Omega_n|$.
(Here $\varepsilon$ and $\Omega$ are the numerical
parameter used in simulations presented below.) One can
minimize the probability of the near-resonant transitions
choosing $\varepsilon=0$.

\section{Errors in creation of an entangled state \\ for remote qubits}
Our matrix approach allows us to estimate the error
in the logic gate (\ref{U}) caused by flips of non-resonant spins
(non-resonant transitions).
Consider a transition between the states $|l\rangle$ and $|l'\rangle$
connected by a flip of the non-resonant $k'$th spin.
The absolute value of the
difference between the $l$th and $l'$th diagonal elements of the
matrix ${\cal H}^{(n)}_{pp'}$
is of the order or greater than $\delta\omega$,
because they belong to different
$2\times 2$ blocks. Since the absolute values of the matrix
elements which connect the different blocks are small,
$|V|\ll\delta\omega$, one can write,
\begin{equation}
\label{psi10}
\psi_{q}=\psi^{(0)}_{q}+
{\sum_{q'}}'{v_{qq'}\over e^{(0)}_q-e^{(0)}_{q'}}
\psi^{(0)}_{q'},
\end{equation}
where prime in the sum means that the term with $q'=q$ is omitted,
$\psi_{q}$ is the eigenfunction of the Hamiltonian ${\cal H}$,
the $q$th eigenstate is related to the $l$th diagonal
element and the $q'$th eigenstate is related to the $l'$th diagonal
element, $v_{qq'}=2V\langle\psi^{(0)}_{q}|I_{k'}^x|\psi^{(0)}_{q'}\rangle$
is the matrix element for the transition between
the states $\psi^{(0)}_{q}$ and $\psi^{(0)}_{q'}$,
the sum over $q'$ takes into consideration all possible
one-spin-flip non-resonant transitions
from the state $|l\rangle$.
Because the matrix ${\cal H}$ is divided into
$2^{N-1}$ relatively independent $2\times 2$ blocks,
the energy, $e^{(0)}_q$ ($e^{(0)}_{q'}$),
and the wave function, $\psi^{(0)}_{q}$ ($\psi^{(0)}_{q'}$),
in Eq. (\ref{psi10}) are, respectively, the eigenvalue and the
eigenfunction with the amplitudes given by Eqs. (\ref{q01}) and
(\ref{q02}) of the effective Hamiltonian,
${\cal H}$, in which  all elements are equal to zero
except the elements related to a single $2\times 2$ block.

The probability of non-resonant transition from the state
$|l\rangle$ to the state $|l'\rangle$ connected
by a flip of the non-resonant $k'$th spin is,
\begin{equation}
\label{psi19}
P_{ll'}=\left|\langle l'|\psi_q\rangle\right|^2.
\end{equation}
Only one term in the sum in Eq. (\ref{psi10})
contributes to the probability $P_{ll'}$.
When the block (\ref{2x2}) is related to the near-resonant
transition ($\Delta\sim J$), then from Eqs. (\ref{q01}) and (\ref{q02})
the eigenfunctions of this block are,
$$
\psi^{(0)}_{q'}\approx
[1-(\Omega^2/32J^2)]|l'\rangle+
(\Omega/4J)|m'\rangle\approx |l'\rangle,
$$
\begin{equation}
\label{appr}
~~~~\psi^{(0)}_{Q'}\approx -(\Omega/4J)|l'\rangle+
[1-(\Omega^2/32J^2)]|m'\rangle\approx |m'\rangle.
\end{equation}
On the other hand, if this block is related to the resonant transition
($\Delta=0$), we have
\begin{equation}
\label{appr1}
\psi^{(0)}_{q'}=(1/\sqrt 2)(|l'\rangle+|m'\rangle), \qquad
\psi^{(0)}_{Q'}=(1/\sqrt 2)(|l'\rangle-|m'\rangle).
\end{equation}
In both cases $v_{qq'}\approx V$, so that
\begin{equation}
\label{psi9}
P_{ll'}\approx
\left({V\over {\cal E}_l-{\cal E}_{l'}}\right)^2\approx
\left({V\over |k-k'|\delta\omega}\right)^2,
\end{equation}
where we put $e^{(0)}_q\approx {\cal E}_l$,
$e^{(0)}_{q'}\approx {\cal E}_{l'}$;
$|k-k'|$ is the ``distance'' from the non-resonant $k'$th
spin to the resonant $k$th spin.

The total probability,
$\mu_{N-1}$ (here the subscript of $\mu$ stands for the number of the
resonant spin),
of generation of all unwanted states by the first $\pi/2$ pulse
in the result of non-resonant transitions is,
\begin{equation}
\label{Pnr10}
\mu_{N-1}=\mu
\sum_{k'=0}^{N-2}\frac 1{|N-1-k'|^2},\qquad
\mu=\left(\Omega\over 2\delta\omega\right)^2.
\end{equation}
The probability of error after applying $2N-2$ pulses is,
\begin{equation}
\label{Pnr}
P=1-\frac 12\prod_{n=1}^{2L-2}(1-\mu_n)-
\frac 12\prod_{n=1}^{2L-2}(1-\mu_n-\varepsilon_n),
\end{equation}
where $\varepsilon_1=0$ (operation (\ref{U0})) and the last two terms
in the right-hand side of
Eq. (\ref{Pnr}) are related to the last two terms
in the right-hand side of Eq. (\ref{U}).

\section{Improved perturbation theory}
\label{sec:improved}
In the consideration presented above, we used the approximate
solutions (\ref{appr}) and (\ref{appr1})
for the wave functions (\ref{q01}) and (\ref{q02}).
A more advanced approach, which also does not require a diagonalization
of large matrices, we use the explicit forms (\ref{q01})
and (\ref{q02}) to express
the wave functions, $\psi^{(0)}_{q'}$ and $\psi^{(0)}_{Q'}$,
of the $2\times 2$ blocks
in Eq. (\ref{psi10}),
\begin{equation}
\label{psi100}
\psi^{(0)}_{q'}=A_{m'}^{q'\,(0)}|m'\rangle
+A_{l'}^{q'\,(0)}|l'\rangle,\qquad
\psi^{(0)}_{Q'}=A_{m'}^{Q'\,(0)}|m'\rangle
+A_{l'}^{Q'\,(0)}|l'\rangle.
\end{equation}
Then, we put the functions, $\psi^{(0)}_{q'}$ and $\psi^{(0)}_{Q'}$,
into Eq. (\ref{psi10}), and
obtain the coefficients, $A_{l}^{q}$, for the wave function
$\psi_{q}$,
\begin{equation}
\label{Amq}
\psi_{q}=\sum_m A_{m}^{q}|m\rangle,
\end{equation}
where the sum in the right-hand side contains $2L$ terms.
Using the functions, $A_{m}^{q}$, we solve
the dynamical equations (\ref{dynamics}) with the energies,
$e_{q}$, computed up to the second order of our perturbation theory, 
\begin{equation}
\label{emq}
e_q=e^{(0)}_q+{\sum_{q'}}'{|v_{qq'}|^2\over e^{(0)}_q-e^{(0)}_{q'}},
\end{equation}
where the prime in the sum means that the term with $q=q'$ is omitted,
$e^{(0)}_q$ is defined by Eqs. (\ref{q01}) or (\ref{q02}).

We call the described above approach the ``improved''
perturbation theory to indicate the difference from the
approach considered in the previous section.
In this approximation each eigenfunction, $\psi_q$,
of the Hamiltonian, $\cal H$, is expanded over $2L$ (see Eq. (\ref{Amq}))
basis functions, $|m\rangle$, with all other coefficients, $A_{m'}^q$, being
equal to zero. Here we use all possible transitions between different
$2\times 2$ blocks which include the two-spin-flip transitions:
a flip of the resonant spin and a flip of a non-resonant spin.
The number of non-zero coefficients in this approximation
is $2^L\times 2L$. It still can be large for large $L$ and can
require large computer memory for simulation. As will be
shown below, under the condition, $\Omega\ll J\ll\delta\omega$, this
approach (the ``improved'' perturbation theory)
gives the results which practically coincide with the exact solution.

\section{Numerical results}

All frequencies in this section are
measured in units of the Ising interaction constant, $J$.
Assume that we are able to correct the errors  with the
probability less than $P_0=10^{-5}$.
Our perturbation theory allows us to calculate the region of
parameters for which
the probability of error, $P$, in realization of the logic gate (\ref{U})
is less than $P_0$. In Figs.~1~(a) and~(b) we plot the diagrams
obtained by solution of Eq.~(\ref{Pnr}) and using the
improved perturbation theory.
In the hatched areas the
probability of generation of unwanted states
is less than $P_0$. One can see that two approaches yield similar
results. They become practically identical at large values of the distance between the neighboring qubits, 
$\delta\omega$.

In almost all quantum algorithms the phase of the wave function
is important. We numerically compared the phase of the wave function on the
boundaries of the hatched
regions in Figs.~1(a,b) with the phase in the centers of these regions,
where $\Omega$ satisfies the $2\pi k$-method, 
and the expression for the phase can be obtained analytically \cite{1}.
The deviation in phase is only $\sim 0.15\%$.
This is much less that the corresponding change in the probability, $P$,
of errors (by several orders).

\begin{figure}[t]
\mbox{\hspace{-5mm}\psfig{file=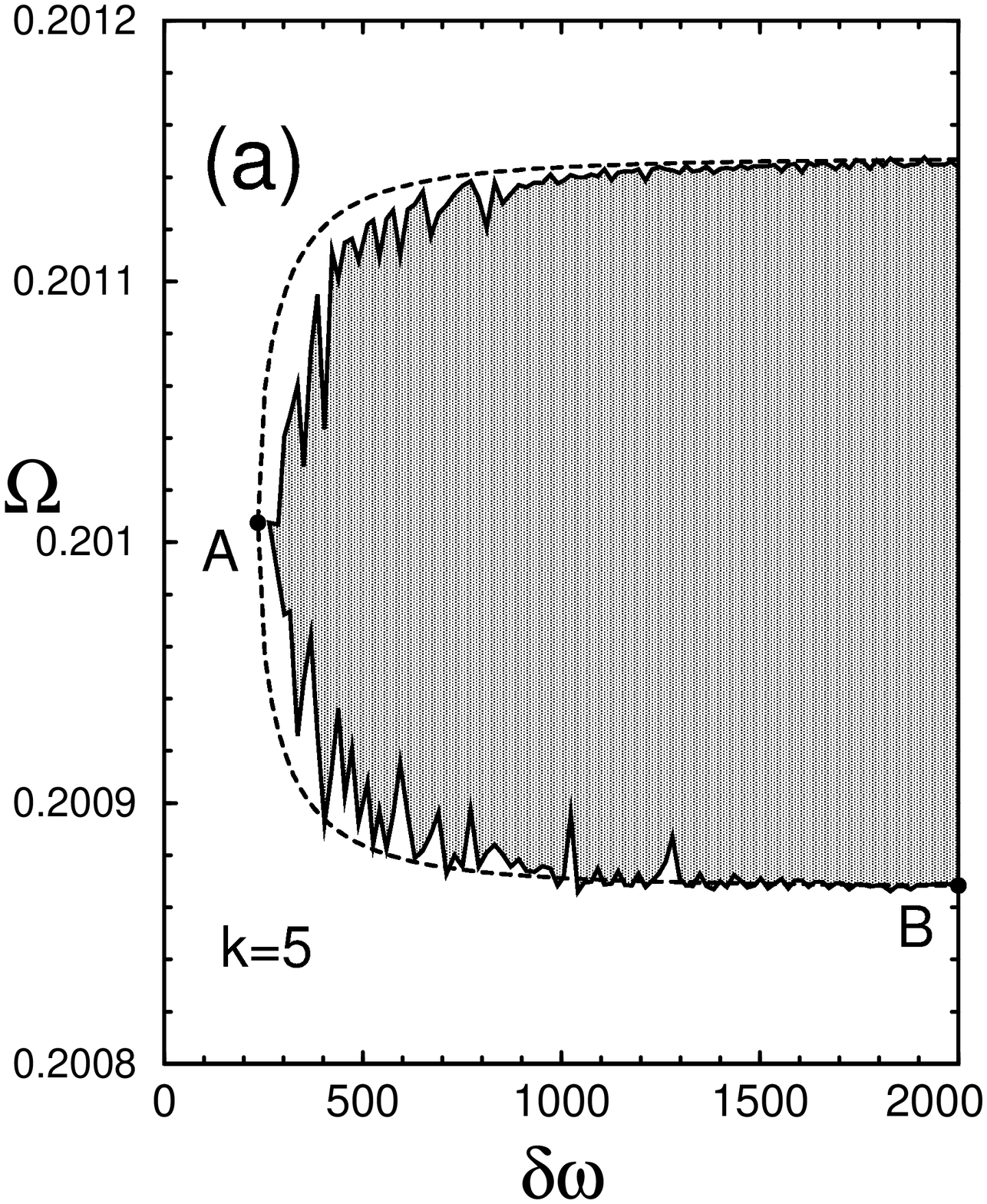,width=8.5cm,height=9cm}
\hspace{-4mm}\psfig{file=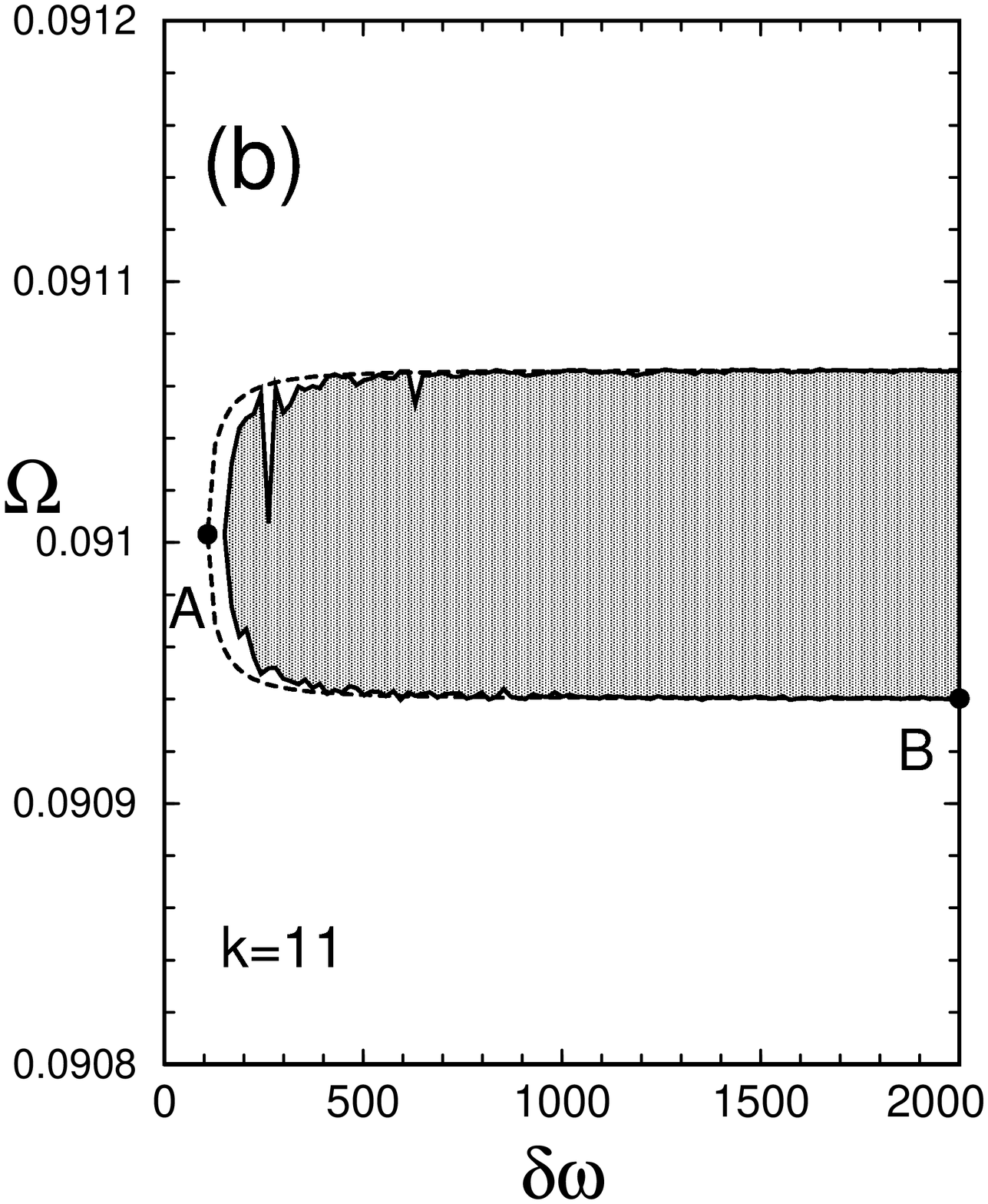,width=8.5cm,height=9cm}}
\caption{The probability of generation of unwanted states, $P$,
at different values of $\delta\omega$ and $\Omega$. In the hatched
regions $P<P_0$ ($P_0=10^{-5}$). The region delimited by the
dashed line is obtained using Eq. (\ref{Pnr}).
The hatched region delimited by the
solid line is obtained using the improved perturbation theory
described in Section {\ref{sec:improved}}. The position of the point A
in $\Omega$ satisfies the $2\pi k$-method (\ref{2pi_k}),
$\Omega=\Omega^{(k)}$, where the values of $k$ are indicated in the figures.
$L=10$.}
\label{fig:1}
\end{figure}

We now analyze the probability of errors as a function of $\delta\omega$.
When the value of
$\delta\omega$ is large enough, the probability of error
(and the widths of the hatched areas in $\Omega$)
becomes practically independent of $\delta\omega$.
This is because at $\delta\omega\gg 1$ and at $\varepsilon\gg\mu$
the error is provided mostly
 by $\varepsilon$, which is independent of $\delta\omega$.
As a consequence, one can, for example, estimate
the widths of the hatched areas at $\delta\omega\gg 1$
taking into account only the near-resonant
transitions. To do this we put in Eq. (\ref{Pnr}) the value $\mu_n=0$
for all $n$ and obtain
\begin{equation}
\label{Pnr1}
P_B=\frac 12\left(1-\prod_{n=1}^{2L-3}(1-\varepsilon_n)\right)=
\frac 12\left(1-(1-\varepsilon)^{2L-3}\right)\approx
\frac {2L-3}2\varepsilon,
\end{equation}
where $\varepsilon_n=\varepsilon\ll 1$, for all $n$.
The positions of the boundaries in $\Omega$ in
Figs. 1~(a,b) can be obtained from the equation $P_B=P_0$, where
$P_B$ is given by Eq. (\ref{Pnr1}) and $\varepsilon$ is a function of
$\Omega$ (see Eq. (\ref{epsilon})).

\begin{figure}[t]
\centerline{\psfig{file=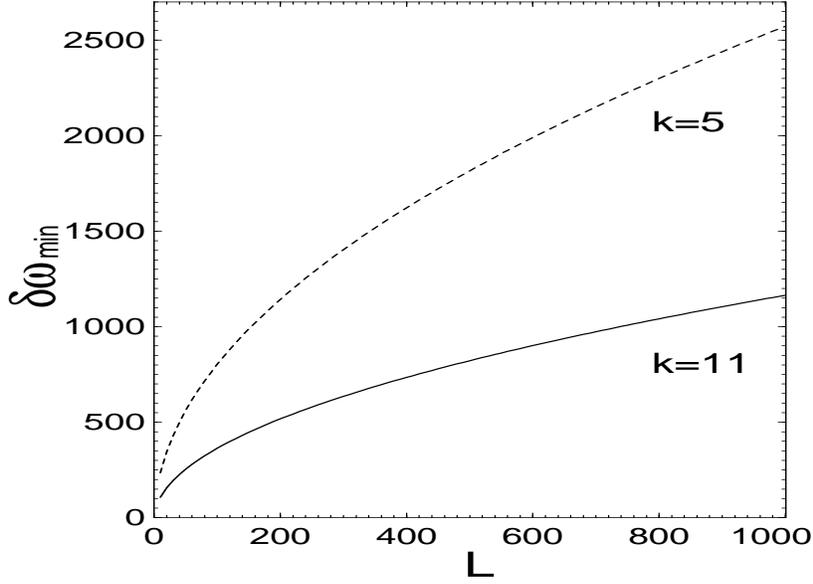,width=11cm,height=8cm}}
\vspace{4mm}
\caption{The minimum value of $\delta\omega$,
$\delta\omega_{min}=\delta\omega_A$ (see Figs. 1 (a,b) for $L=10$),
required to
make the error in the logic gate (\ref{U}) below the
threshold $P_0=10^{-5}$, as a function of number of qubits, $L$.
The value of $\Omega=\Omega^{(k)}$ satisfies the conditions of
the $2\pi k$-method. The values of $k$ are indicated in the figure.}
\label{fig:2}
\end{figure}

From Figs. 1~(a)~and~(b) one can see that even when the
condition of $2\pi k$-method is satisfied, the error can be large
when $\delta\omega$ (or the gradient of the magnetic field $B^z(z)$)
is relatively small. Thus, at $\delta\omega<\delta\omega_A$,
where $\delta\omega_A$ is the coordinate of the point A in $\delta\omega$
in Figs. 1~(a,b), the error
is more than $P_0$. Since the position of the point A in $\Omega$
satisfies the $2\pi k$-method,
this indicates that even in the case
when the near-resonant processes are suppressed by the $2\pi k$ method
($\varepsilon=0$) the non-resonant transitions
can make the error larger than the threshold, $P_0$. We can
not suppress entirely the non-resonant transitions, defined by
the values of $\mu=(\Omega/2\omega)^2$ and $L$, like we did with the
near-resonant transitions. The value of $\Omega$ cannot be
decreased considerably because decreasing of $\Omega$ makes
the quantum computer very slow, so that the quantum state can be
destroyed by decoherence due to possible influence of environment.
Hence, one can make the value of $\mu$ small by
increasing $\delta\omega$.
In Fig.~2 we plot the minimum value of $\delta\omega=\delta\omega_A$ as a
function of the number of qubits, $L$, which was computed using Eq.~(\ref{Pnr}).

\begin{figure}[t]
\centerline{\psfig{file=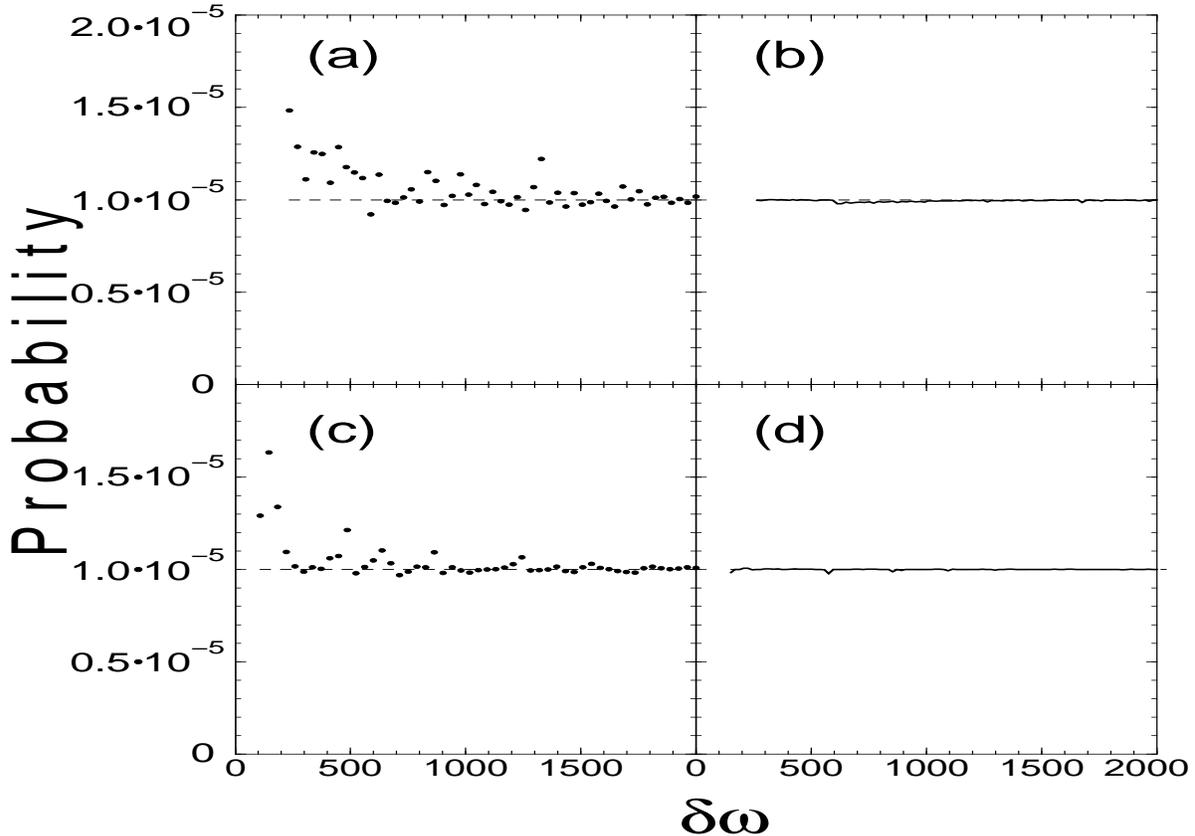,width=16cm,height=11.5cm}}
\caption{The exact solution for the
probability of generation of unwanted states, $P$, computed using
the parameters which correspond to
(a) dashed curve AB in Fig. 1 (a) (obtained using Eq. (\ref{Pnr})),
(b) lower boundary of the hatched region in Fig. 1 (a)
(obtained using the improved perturbation theory),
(c) dashed curve AB in Fig. 1 (b) (obtained using Eq. (\ref{Pnr})),
(d) lower boundary of the hatched region in Fig. 1 (b)
(obtained using the improved perturbation theory).
The dashed lines indicate the solutions obtained using
the corresponding perturbation theory for the same parameters.}
\label{fig:3}
\end{figure}

One can see that $\delta\omega_A$ becomes large for large $L$. Thus,
for example, for a protons with $J/(2\pi)\sim 100$ Hz
(for estimations in this paragraph we use the dimensional units)
with the distance between the neighboring spins $a=2$ nm, the value
$\delta\omega/J=1000$ yield the gradient of the magnetic field
$\delta\omega/(\gamma a\cos\theta)\sim 2\times 10^6$ T/m.
From Fig.~2 one can see that this is the
minimum value of the gradient of the magnetic field for
$L_{max}\approx 155$ when $\Omega=\Omega^{(5)}$ and
$L_{max}\approx 740$ when $\Omega=\Omega^{(11)}$
required to make the error less than $P_0=10^5$.
At $L>L_{max}$, at a given gradient of the magnetic field,
and at $\Omega\approx\Omega^{(k)}$, $k=5$ or $11$,
the error will be always larger than $P_0$.

In Figs.~3~(a,b) we test our perturbation theory by using the exact
numerical solution obtained by a diagonalization of $2^L\times 2^L$
matrices and using Eq. (\ref{dynamics}). One can see that there is
good correspondence with the exact numerical solution
for the results obtained using Eq. (\ref{Pnr}),
and practically exact correspondence for the solution obtained using
the improved perturbation theory. The similar correspondence can be
demonstrated for other parameters ($\delta\omega,\,\Omega$).

\section{Conclusion}
We developed the perturbation theory which allows us
to estimate the errors in implementation of the quantum
logic gates by the radio-frequency pulses in the solid-state
system with large number (1000 and more) of qubits.
Our perturbation approach correctly describes
the behavior of the quantum system in the large Hilbert space
(the Hilbert space with large number of states) and predicts the
final quantum state of the system after action of the sequence of
pulses with different frequencies. This is possible because
in the system there are small parameters, characterizing the
probabilities $\varepsilon$, of the near-resonant transitions, and
probabilities, $\mu$, of the non-resonant transitions, 
which are small, $\varepsilon\ll 1$ and $\mu\ll 1$,
when the conditions $\Omega\ll J\ll\delta\omega$ are satisfied.
Our approach allows one to control the quantum logic operations in the
system with large number of qubits and to minimize the error
caused by the internal decoherence (non-resonant processes).

\section{Acknowledgments}
The paper was supported by the Department of Energy (DOE) under
contract W-7405-ENG-36, by the National Security Agency (NSA), and by the
Advanced Research and Development Activity (ARDA).

{}

\end{document}